\documentclass{article}

\usepackage{graphicx} 
\usepackage{subfigure}

\usepackage{pstricks}
\usepackage{tikz}

\usepackage{natbib}

\usepackage{algorithm}
\usepackage[noend]{algorithmic}

\usepackage{hyperref}

\usepackage[accepted]{icml2013}

\usepackage{amsmath}
\usepackage{mathrsfs}
\usepackage{amsthm}
\usepackage{amssymb}

\newtheorem{defn}{Definition}

\icmltitlerunning{Tree-Independent Dual-Tree Algorithms}

\begin{document}

\twocolumn[
\vspace*{-1.8em}
\icmltitle{Tree-Independent Dual-Tree Algorithms}

\vspace*{-1.2em}
\icmlauthor{Ryan R. Curtin}{rcurtin@cc.gatech.edu}
\icmlauthor{William B. March}{march@gatech.edu}
\icmlauthor{Parikshit Ram}{p.ram@gatech.edu}
\icmlauthor{David V. Anderson}{anderson@gatech.edu}
\icmlauthor{Alexander G. Gray}{agray@cc.gatech.edu}
\icmlauthor{Charles L. Isbell, Jr.}{isbell@cc.gatech.edu}
\icmladdress{Georgia Institute of Technology,
            266 Ferst Drive NW, Atlanta, GA 30332 USA}

\icmlkeywords{some keywords}

\vskip 0.05in
]

\begin{abstract}
\vspace*{-0.1em}
{\em Dual-tree algorithms} are a widely used class of branch-and-bound
algorithms.  Unfortunately, developing dual-tree algorithms for use with
different trees and problems is often complex and burdensome.  We
introduce a four-part logical split: the tree, the traversal,
the point-to-point base case, and the pruning rule.  We provide a meta-algorithm
which allows development of dual-tree algorithms in a tree-independent manner
and easy extension to entirely new types of trees.  Representations are provided
for five common algorithms; for $k$-nearest neighbor search, this
leads to a novel, tighter pruning bound. The meta-algorithm also allows
straightforward extensions to massively parallel settings.
\end{abstract}

\vspace*{-2.2em}
\section{Introduction}
\label{sec:intro}
\vspace*{-0.2em}


In large-scale machine learning applications, algorithmic scalability is
paramount.  Hence, much study has been put into fast algorithms for machine
learning tasks.  One commonly-used approach is to build trees on data and then
use branch-and-bound algorithms to minimize runtime.  A popular example of a
branch-and-bound algorithm is the use of the trees for nearest neighbor search,
pioneered by Bentley \yrcite{bentley1975}, and subsequently modified to use two
trees (``dual-tree'') \cite{nbody}.  Later, an optimized tree structure,
the cover tree, was designed \cite{langford2006}, giving provably
linear scaling in the number of queries \cite{ram2009}---a significant
improvement over the quadratically-scaling brute-force algorithm.

Asymptotic speed gains as dramatic as described above are common for dual-tree
branch-and-bound algorithms.  These types of algorithms can be applied to a
class of problems referred to as `$n$-body problems' \cite{nbody}.  The
$n$-point correlation, important in astrophysics, is an $n$-body problem and can
be solved quickly with trees \cite{march2012}.  In addition, Euclidean minimum
spanning trees can be found quickly using tree-based algorithms
\cite{march2010}.  Other dual-tree algorithms include kernel density estimation
\cite{gray2003}, mean shift \cite{wang2007}, Gaussian summation \cite{lee2006},
kernel density estimation, fast singular value decomposition
\cite{holmes2008quic}, range search, furthest-neighbor search, and many others.


The dual-tree algorithms referenced above are each quite similar, but no formal
connections between the algorithms have been established.  The
types of trees used to solve each problem may differ, and in addition, the
manner in which the trees are traversed can differ (depending on the
problem or the tree).  In practice, a researcher may have to
implement entirely separate algorithms to solve the same problems with different
trees; this is time-consuming and makes it difficult to explore the
properties of tree types that make them more suited for particular problems.
Worse yet, parallel dual-tree algorithms are difficult to
develop and appear to be far more complex than serial implementations; yet, both
solve the same problem.  We make these contributions to address these
shortcomings:


\begin{itemize}
  \vspace*{-1em}
  \item A {\bf representation} of dual-tree algorithms as {\bf four separate
components}: a space tree, a traversal, a base case, and a pruning rule.
  \vspace*{-0.5em}
  \item A {\bf meta-algorithm} that produces dual-tree algorithms, given those
four separate components.
  \vspace*{-1.5em}
  \item Base cases and pruning rules for {\bf a variety of dual-tree
algorithms}, which can be used with \textbf{any} space tree and \textbf{any}
traversal.
  \vspace*{-0.5em}
  \item A {\bf theoretical framework}, used to prove the correctness of these
meta-algorithms and develop a new, tighter bound for $k$-nearest neighbor
search.
  \vspace*{-0.5em}

  \item Implications of our representation, including {\bf easy creation of
large-scale distributed dual-tree algorithms} via our meta-algorithm.

\end{itemize}


\section{Overview of Meta-Algorithm}
\label{sec:overview}

In other works, dual-tree algorithms are described as standalone algorithms that
operate on a query dataset $S_q$ and a reference dataset $S_r$.  By observing
commonalities in these algorithms, we propose the following logical split of any
dual-tree algorithm into four parts:

\vspace*{-0.7em}
\begin{itemize}
  \item A \textit{space tree} (a type of data structure).

  \item A \textit{pruning dual-tree traversal}, which visits nodes in two space
trees, and is parameterized by a \texttt{BaseCase()} and a \texttt{Score()}
function.

  \item A \texttt{BaseCase()} function that defines the action to take on a
combination of points.

  \item A \texttt{Score()} function that determines if a subtree should be
visited during the traversal.
\end{itemize}
\vspace*{-0.7em}

We can use this to define a {\bf meta-algorithm}:

\vspace*{-0.7em}
\begin{quote}
\textit{Given a type of space tree, a pruning dual-tree traversal, a
}\texttt{BaseCase()}\textit{ function, and a }\texttt{Score()}\textit{ function,
use the pruning dual-tree traversal with the given }\texttt{BaseCase()}\textit{
and }\texttt{Score()}\textit{ functions on two space trees $\mathscr{T}_q$
(built on $S_q$) and $\mathscr{T}_r$ (built on $S_r$).}
\end{quote}
\vspace*{-0.7em}

In Sections \ref{sec:trees} and \ref{sec:traversers}, space trees, traversals,
and related quantities are rigorously defined.  Then, Sections
\ref{sec:knn}--\ref{sec:kde} define \texttt{BaseCase()} and \texttt{Score()}
for various dual-tree algorithms.  Section \ref{sec:discussion} discusses
implications and future possibilities, including large-scale parallelism.

\section{Space Trees}
\label{sec:trees}

To develop a framework for understanding dual-tree algorithms, we must
introduce some terminology.

\begin{defn}
\label{def:spacetree}
A \textbf{space tree} on a dataset $S \in \Re^{N \times D}$ is an undirected,
connected, acyclic, rooted simple graph with the following properties:

\vspace*{-0.5em}
\begin{itemize}
  \item Each \textit{node} (or vertex), holds a number of points (possibly
zero) and is connected to one parent node and a number of child nodes (possibly
zero).
  \item There is one node in every space tree with no parent; this
is the \textit{root node} of the tree.
  \item Each point in $S$ is contained in at least one node of the
tree.
  \item Each node $\mathscr{N}$ of the tree has a convex subset of
$\Re^{D}$ that contains each of the points in that node as well as the convex
subsets represented by each child of the node.
\end{itemize}
\end{defn}

\vspace*{-0.6em}
Notationally, we use the following conventions:
\vspace*{-0.5em}

\begin{itemize}
  \item The set of child nodes of a node $\mathscr{N}_i$ is denoted
$\mathscr{C}(\mathscr{N}_i)$ or $\mathscr{C}_i$.
  \item The set of points held in a node $\mathscr{N}_i$ is denoted
$\mathscr{P}(\mathscr{N}_i)$ or $\mathscr{P}_i$.
  \item The convex subset represented by node $\mathscr{N}_i$ is denoted
$\mathscr{S}(\mathscr{N}_i)$ or $\mathscr{S}_i$.
  \item The set of descendant nodes of a node $\mathscr{N}_i$, denoted
$\mathscr{D}^n(\mathscr{N}_i)$ or $\mathscr{D}^n_i$, is the set of nodes
$\mathscr{C}(\mathscr{N}_i) \:\cup\: \mathscr{C}(\mathscr{C}(\mathscr{N}_i))
\:\cup\: \dots$ .
  \item The set of descendant points of a node $\mathscr{N}_i$, denoted
$\mathscr{D}^p(\mathscr{N}_i)$ or $\mathscr{D}^p_i$, is the set of points $\{ \:
p : p \in \mathscr{D}^n(\mathscr{N}_i) \cup \mathscr{P}(\mathscr{N}_i) \: \}$.
  \item The parent of a node $\mathscr{N}_i$ is denoted
$\operatorname{Par}(\mathscr{N}_i)$.
\end{itemize}

\vspace*{-0.6em}

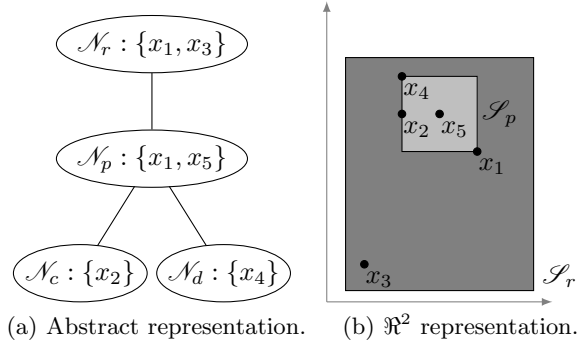
\begin{figure}[t!]
  \centering
  \subfigure[Abstract representation.]{
    \begin{tikzpicture}[>=latex,line join=bevel,scale=0.6]
  \pgfsetlinewidth{1bp}
\pgfsetcolor{black}
  \draw [thin] (70.107bp,72.571bp) .. controls (64.745bp,63.992bp) and (58.173bp,53.476bp)  .. (46.793bp,35.269bp);
  \draw [thin] (81bp,143.83bp) .. controls (81bp,136.13bp) and (81bp,126.97bp)  .. (81bp,108.41bp);
  \draw [thin] (91.893bp,72.571bp) .. controls (97.255bp,63.992bp) and (103.83bp,53.476bp)  .. (115.21bp,35.269bp);
\begin{scope}
  \definecolor{strokecol}{rgb}{0.0,0.0,0.0};
  \pgfsetstrokecolor{strokecol}
  \draw [thin] (126bp,18bp) ellipse (42bp and 18bp);
  \draw (126bp,18bp) node {$\mathscr{N}_d: \{ x_4 \}$};
\end{scope}
\begin{scope}
  \definecolor{strokecol}{rgb}{0.0,0.0,0.0};
  \pgfsetstrokecolor{strokecol}
  \draw [thin] (36bp,18bp) ellipse (42bp and 18bp);
  \draw (36bp,18bp) node {$\mathscr{N}_c: \{ x_2 \}$};
\end{scope}
\begin{scope}
  \definecolor{strokecol}{rgb}{0.0,0.0,0.0};
  \pgfsetstrokecolor{strokecol}
  \draw [thin] (81bp,162bp) ellipse (60bp and 18bp);
  \draw (81bp,162bp) node {$\mathscr{N}_r: \{ x_1, x_3 \}$};
\end{scope}
\begin{scope}
  \definecolor{strokecol}{rgb}{0.0,0.0,0.0};
  \pgfsetstrokecolor{strokecol}
  \draw [thin] (81bp,90bp) ellipse (60bp and 18bp);
  \draw (81bp,90bp) node {$\mathscr{N}_p: \{ x_1, x_5 \}$};
\end{scope}
\end{tikzpicture}
    \label{fig:sptree_abstract}
  }
  \subfigure[$\Re^2$ representation.]{
    \begin{tikzpicture}
  \coordinate (Origin) at (0, 0);
  \coordinate (XAxisMin) at (0, 0);
  \coordinate (XAxisMax) at (3, 0);
  \coordinate (YAxisMin) at (0, 0);
  \coordinate (YAxisMax) at (0, 4);
  \draw [thin, gray, -latex] (XAxisMin) -- (XAxisMax);
  \draw [thin, gray, -latex] (YAxisMin) -- (YAxisMax);

  \filldraw[fill=gray] (0.25, 0.15) -- (0.25, 3.25) -- (2.75, 3.25) -- (2.75, 0.15) -- cycle;
  \node [ ] at (3.1, 0.35) { $\mathscr{S}_r$ };

  \filldraw[fill=lightgray] (1, 2) -- (1, 3) -- (2, 3) -- (2, 2) -- cycle;
  \node [ ] at (2.3, 2.5) { $\mathscr{S}_p$ };

  \node [draw, circle, inner sep=1pt, fill] at (2, 2) { };
  \node [ ] at (2.2, 1.8) { $x_1$ };

  \node [draw, circle, inner sep=1pt, fill] at (1, 2.5) { };
  \node [ ] at (1.2, 2.3) { $x_2$ };

  \node [draw, circle, inner sep=1pt, fill] at (0.5, 0.5) { };
  \node [ ] at (0.7, 0.3) { $x_3$ };

  \node [draw, circle, inner sep=1pt, fill] at (1, 3) { };
  \node [ ] at (1.2, 2.8) { $x_4$ };

  \node [draw, circle, inner sep=1pt, fill] at (1.5, 2.5) { };
  \node [ ] at (1.7, 2.3) { $x_5$ };

\end{tikzpicture}
    \label{fig:sptree_r2}
  }
  \vspace*{-1em}
  \caption{An example space tree.}
  \label{fig:sptree}
  \vspace*{-0.7em}
\end{figure}

An abstract representation of an example space tree on a five-point dataset in
$\Re^{2}$ is shown in Figure~\ref{fig:sptree_abstract}.  In this illustration,
$\mathscr{N}_r$ is the root node of the tree; it has no parent and it contains
the points $x_3$ and $x_1$ (that is, $\mathscr{P}_r = \{ x_1, x_5 \}$.  The node
$\mathscr{N}_p$ contains points $x_1$ and $x_5$ and has children $\mathscr{N}_c$
and $\mathscr{N}_d$ (which each have no children and contain points $x_2$ and
$x_4$, respectively).  In Figure~\ref{fig:sptree_r2}, the points in the tree and
the subsets $\mathscr{S}_r$ (darker rectangle) and $\mathscr{S}_p$ (lighter
rectangle) are plotted. $\mathscr{S}_c = \{ x_2 \}$ and $\mathscr{S}_d = \{ x_4
\}$ are not labeled.

\vspace*{0.5em}
\begin{defn}
The \textbf{minimum distance} between two nodes $\mathscr{N}_i$ and
$\mathscr{N}_j$ is defined as

\vspace*{-1.8em}
$$d_{min}(\mathscr{N}_i, \mathscr{N}_j) = \min \left\{ \: \| p_i - p_j \| \;
\forall \; p_i \in \mathscr{D}^p_i, p_j \in \mathscr{D}^p_j \: \right\} .$$
\end{defn}

\begin{defn}
The \textbf{maximum distance} between two nodes $\mathscr{N}_i$ and
$\mathscr{N}_j$ is defined as

\vspace*{-1.8em}
$$d_{max}(\mathscr{N}_i, \mathscr{N}_j) = \max \left\{ \: \| p_i - p_j \| \;
\forall \; p_i \in \mathscr{D}^p_i, p_j \in \mathscr{D}^p_j \: \right\} .$$
\end{defn}

\begin{defn}
The \textbf{maximum child distance} of a node $\mathscr{N}_i$ is defined as the
maximum distance between the centroid $C_i$ of $\mathscr{S}_i$ and each point in
$\mathscr{P}_i$:

\vspace*{-1.2em}
$$\rho(\mathscr{N}_i) = \max_{p \in \mathscr{P}_i} \| C_i - p \|.$$
\end{defn}

\begin{defn}
The \textbf{maximum descendant distance} of a node $\mathscr{N}_i$ is defined as
the maximum distance between the centroid $C_i$ of $\mathscr{S}_i$ and points in $\mathscr{D}^p_i$:

\vspace*{-1.3em}
$$\lambda(\mathscr{N}_i) = \max_{p \in \mathscr{D}^p_i} \| C_i - p \|.$$
\end{defn}
\vspace*{-1.0em}



It is straightforward to show that $kd$-trees, octrees, metric trees, ball
trees, cover trees \cite{langford2006}, R-trees, and vantage-point trees all
satisfy the conditions of a space tree.  The quantities $d_{\min}(\mathscr{N}_q,
\mathscr{N}_r)$, $d_{\max}(\mathscr{N}_q, \mathscr{N}_r)$,
$\lambda(\mathscr{N}_i)$, and $\rho(\mathscr{N}_i)$ are easily derived (or
bounded, which in many cases is sufficient) for each of these types of trees.

\vspace*{-0.2em}
For a $kd$-tree, $d_{\min}(\mathscr{N}_i, \mathscr{N}_j)$ is bounded below by
the minimum distance between $\mathscr{S}_i$ and $\mathscr{S}_j$;
$d_{\max}(\mathscr{N}_i, \mathscr{N}_j)$ is bounded above similarly by the
maximum distance between $\mathscr{S}_i$ and $\mathscr{S}_j$.  Both
$\rho(\mathscr{N}_i)$ and $\lambda(\mathscr{N}_i)$ are bounded above by
$d_{\max}(C_i, \mathscr{N}_i)$.


\vspace*{-0.2em}
For the cover tree \cite{langford2006}, each node $\mathscr{N}_i$ contains only
one point $p_i$ and has `scale' $s_i$.  $d_{\min}(\mathscr{N}_i, \mathscr{N}_j)$
is bounded below by $d(p_i, p_j) - 2^{s_i + 1} - 2^{s_j + 1}$ and
$d_{\max}(\mathscr{N}_i, \mathscr{N}_j)$ is bounded above by $d(p_i, p_j) +
2^{s_i + 1} + 2^{s_j + 1}$.  Because $p_i$ is the centroid of $\mathscr{S}_i$,
$\rho(\mathscr{N}_i) = 0$.  $\lambda(\mathscr{N}_i)$ is simply $2^{s_i + 1}$.

\vspace*{-0.5em}
\section{Tree Traversals}
\label{sec:traversers}

In general, the nodes of each space tree can be traversed in a number of ways.
However, there have been no attempts to formalize tree traversal.  Therefore, we
introduce several definitions which will be useful later.

\vspace*{0.4em}
\begin{defn}
A \textbf{single-tree traversal} is a process that, given a space tree, will
visit each node in that tree once and perform a computation on any points
contained within the node that is being visited.
\end{defn}

As an example, the standard depth-first traversal or breadth-first traversal
are single-tree traversals.  From a programming perspective, the computation in
the single-tree traversal can be implemented with a simple callback
\texttt{BaseCase(point)} function.  This allows the computation to be entirely
independent of the single-tree traversal itself.  As an example, a simple
single-tree algorithm to count the number of points in a given tree would
increment a counter variable each time \texttt{BaseCase(point)} was called.
However, this concept by itself is not very useful; without pruning branches, no
computations can be avoided.

\vspace*{0.4em}
\begin{defn}
A \textbf{pruning single-tree traversal} is a process that, given a space tree,
will visit nodes in the tree and perform a computation to assign a score to that
node.  If the score is above some bound, the node is ``pruned'' and none of its
descendants will be visited; otherwise, a computation is performed on any points
contained within that node.  If no nodes are pruned, then the traversal will
visit each node in the tree once.
\end{defn}
\vspace*{-0.3em}

Clearly, a pruning single-tree traversal that does not prune any nodes is a
single-tree traversal.  A pruning single-tree traversal can be implemented with
two callbacks: \texttt{BaseCase(point)} and \texttt{Score(node)}.  This allows
both the point-to-point computation and the scoring to be entirely independent
of the traversal.  Thus, single-tree branch-and-bound algorithms can be
expressed in a tree-independent manner.  Extensions to the dual-tree case are
given below.

\vspace*{0.4em}
\begin{defn}
A \textbf{dual-tree traversal} is a process that, given two space trees
$\mathscr{T}_q$ (query tree) and $\mathscr{T}_r$ (reference tree), will visit
every combination of nodes $(\mathscr{N}_q, \mathscr{N}_r)$ once, where
$\mathscr{N}_q \in \mathscr{T}_q$ and $\mathscr{N}_r \in \mathscr{T}_r$.  At
each visit $(\mathscr{N}_q, \mathscr{N}_r)$, a computation is performed between
each point in $\mathscr{N}_q$ and each point in $\mathscr{N}_r$.
\end{defn}


\vspace*{0.2em}
\begin{defn}
\label{def:dtpt}
A \textbf{pruning dual-tree traversal} is a process which, given two space
trees $\mathscr{T}_q$ (query tree) and $\mathscr{T}_r$ (reference
tree), will visit combinations of nodes $(\mathscr{N}_q, \mathscr{N}_r)$ such
that $\mathscr{N}_q \in \mathscr{T}_q$ and $\mathscr{N}_r \in \mathscr{T}_r$ no
more than once, and perform a computation to assign a score to that combination.
If the score is above some bound, the combination is pruned and no combinations
$(\mathscr{N}_{qc}, \mathscr{N}_{rc})$ such that $\mathscr{N}_{qc} \in
\mathscr{D}^n_q$ and $\mathscr{N}_{rc} \in \mathscr{D}^n_r)$ will be visited;
otherwise, a computation is performed between each point in $\mathscr{N}_q$ and
each point in $\mathscr{N}_r$.
\end{defn}
\vspace*{-0.3em}

Similar to the pruning single-tree traversal, a pruning dual-tree algorithm can
use two callback functions \texttt{BaseCase($p_q$, $p_r$)} and
\texttt{Score($\mathscr{N}_q$, $\mathscr{N}_r$)}.  An example implementation of
a depth-first pruning dual-tree traversal is given in
Algorithm \ref{alg:depth_traversal}.  The traversal is started on the root of
the $\mathscr{T}_q$ and the root of $\mathscr{T}_r$.

Algorithm \ref{alg:depth_traversal} provides only one example of a commonly-used
pruning dual-tree traversal.  Other possibilities not explicitly documented here
include breadth-first traversals and the unique cover tree dual-tree traversal
described by Beygelzimer et al. \yrcite{langford2006}, which can be adapted to
the generic space tree case.

\begin{algorithm}[tb]
\begin{algorithmic}
    \STATE \textbf{if} \texttt{Score($\mathscr{N}_q$, $\mathscr{N}_r$) $=
\infty$} \textbf{then return}
    \medskip
    \FORALL{$p_q \in \mathscr{P}_q, p_r \in \mathscr{P}_r$}
      \STATE \texttt{BaseCase($p_q$, $p_r$)}
    \ENDFOR
    \medskip
    \FORALL{$\mathscr{N}_{qc} \in \mathscr{C}_q, \mathscr{N}_{rc} \in
\mathscr{C}_r$}
      \STATE \texttt{DepthFirstTraversal($\mathscr{N}_{qc}$,
$\mathscr{N}_{rc}$)}
    \ENDFOR
  \end{algorithmic}

  \caption{\texttt{DepthFirstTraversal($\mathscr{N}_q$, $\mathscr{N}_r$)}.}
  \label{alg:depth_traversal}
\end{algorithm}

The rest of this work is devoted to using these concepts to represent existing
dual-tree algorithms in the four parts described in Section \ref{sec:overview}.



\section{$k$-Nearest Neighbor Search}
\label{sec:knn}


$k$-nearest neighbor search is a well-studied problem with a plethora of
algorithms and results.  The problem can be stated as follows:

Given a query dataset $S_q \in \Re^{n \times d}$, a reference dataset $S_r \in
\Re^{m \times d}$, and an integer $k : 0 < k < m$, for each point $p_q \in S_q$,
find the $k$ nearest neighbors in $S_r$ and their distances from $p_q$.  The
list of nearest neighbors for a point $p_q$ can be referred to as $N_{p_q}$ and
the distances to nearest neighbors for $p_q$ can be referred to as $D_{p_q}$.
Thus, the $k$-th nearest neighbor to point $p_q$ is $N_{p_q}[k]$ and $D_{p_q}[k]
= \| p_q - N_{p_q}[k] \|$.

This can be solved using a brute-force approach: compare every possible point
combination and store the $k$ smallest distance results for each $p_q$.
However, this scales poorly -- $O(nm)$; hence the importance of fast algorithms
to solve the problem.  Many existing algorithms employ tree-based
branch-and-bound strategies \cite{langford2006, cover1967, friedman1977,
fukunaga1975, nbody, ram2009}.

We unify all of these branch-and-bound strategies by defining methods
\texttt{BaseCase($p_q$, $p_r$)} and \texttt{Score($\mathscr{N}_q$,
$\mathscr{N}_r$)} for use with a pruning dual-tree traversal.

At the initialization of the tree traversal, the lists $N_{p_q}$ and $D_{p_q}$
are empty lists for each query point $p_q$.  After the traversal is complete,
for a query point $p_q$, the set $\{ N_{p_q}[1] , ... , N_{p_q}[k] \}$ is the
ordered set of $k$ nearest neighbors of the point $p_q$, and each $ D_{p_q}[i] =
\| p_q - N_{p_q}[i] \| $.  If we assume that $D_{p_q}[i] = \infty$ if $i$ is
greater than the length of $D_{p_q}$, we can formulate \texttt{BaseCase()} as
given in Algorithm \ref{alg:knn_base_case}\footnote{
In practice, $k$-nearest-neighbors is often run with identical reference and
query sets.  In that situation it may be useful to modify this implementation of
\texttt{BaseCase()} so that a point does not return itself as the nearest
neighbor (with distance 0).}.

With the base case established, only the pruning rule remains.  A valid pruning
rule will, for a given query node $\mathscr{N}_q$ and reference node
$\mathscr{N}_r$, prune the reference subtree rooted at $\mathscr{N}_r$ if
and only if it is known that there are no points in $\mathscr{D}^p_r$ that are
in the set of $k$ nearest neighbors of any points in $\mathscr{D}^p_q$.
Thus, at any point in the traversal, we can prune the combination
$(\mathscr{N}_q, \mathscr{N}_r)$ if and only if $ d_{\min}(\mathscr{N}_q,
\mathscr{N}_r) \ge B_1(\mathscr{N}_q) $, where

\vspace*{-2em}
$$ B_1(\mathscr{N}_q) = \max_{p \in \mathscr{D}_q^p} D_p[k] . $$
\vspace*{-2em}

Now, we can describe this bound recursively.  This is important for
implementation; a recursive function can cache previous calculations for large
speedups.

\vspace*{-1em}
\begin{eqnarray*}
B_1(\mathscr{N}_q) &=& \max \big\{ \max_{p \in
\mathscr{P}_q} D_p[k] , \max_{p \in \mathscr{D}_q^p, p \not\in
\mathscr{P}_q } D_p[k] \big\} \\
 &=& \max \big\{ \max_{p \in \mathscr{P}_q} D_p[k], \max_{\mathscr{N}_c
\in \mathscr{C}_q} \{ \max_{p \in \mathscr{D}^p_c} D_p[k] \} \big\} \\
 &=& \max \big\{ \max_{p \in \mathscr{P}_q} D_p[k] , \max_{\mathscr{N}_c
\in \mathscr{C}_q} B_1(\mathscr{N}_c) \big\} \\
\end{eqnarray*}
\vspace*{-2em}

\begin{algorithm}[tb]
\begin{algorithmic}
    \STATE {\bfseries Input:} query point $p_q$, reference point $p_r$, list of
$k$ nearest candidate points $N_{p_q}$ and $k$ candidate distances $D_{p_q}$
(both ordered by ascending distance)
    \STATE {\bfseries Output:} distance $d$ between $p_q$ and $p_r$

    \medskip
    \STATE $d \gets \| p_q - p_r \|$
    \medskip

    \IF{$d < D_{p_q}[k]$ \AND \texttt{BaseCase($p_q$, $p_r$)} not yet called}
    \STATE  insert $d$ into ordered list $D_{p_q}$ and truncate list to length
$k$
    \STATE  insert $p_r$ into $N_{p_q}$ such that $N_{p_q}$ is ordered by
distance and truncate list to length $k$
    \ENDIF

    \RETURN $d$
  \end{algorithmic}

  \caption{$k$-nearest-neighbors \texttt{BaseCase()}}
  \label{alg:knn_base_case}
\end{algorithm}

Suppose we have, at some point in the traversal, two points $p_0, p_1 \in
\mathscr{D}^p_q$ for some node $\mathscr{N}_q$, with
$D_{p_0}[k] = \infty$ and $D_{p_1}[k] < \infty$.  This means there exist $k$ points $\{ p_r^1, \dots, p_r^k \}$ in
$S_r$ such that $d(p_1, p_r^i) \le D_{p_1}[k]$ for $i = \{ 1, \dots, k \}$.
Because $p_0, p_1 \in \mathscr{D}^p_q$, we can apply the triangle inequality to
see that $d(p_0, p_1) \le 2 \lambda(\mathscr{N}_q)$.  Therefore, $d(p_0, p_r^i)
\le D_{p_1}[k] + 2 \lambda(\mathscr{N}_q)$ for $i = \{ 1, \dots, k \}$. Using
this observation we can construct an alternate bound function
$B_2(\mathscr{N}_q)$:

$$ B_2(\mathscr{N}_q) = \min_{p \in \mathscr{D}^p_q} D_p[k] + 2
\lambda(\mathscr{N}_q) $$

which can, like $B_1(\mathscr{N}_q)$, be rearranged to provide a recursive
definition.  In addition, if $p_0 \in \mathscr{P}_q$ and $p_1 \in
\mathscr{D}^p_q$, we can bound $d(p_0, p_1)$ more tightly with
$\rho(\mathscr{N}_q) + \lambda(\mathscr{N}_q)$ instead of $2
\lambda(\mathscr{N}_q)$.  These observations yield

\begin{equation*}
\begin{split}
B_2(\mathscr{N}_q) = \min &\left\{ \min_{p \in \mathscr{P}_q} (D_p[k] +
\rho(\mathscr{N}_q) + \lambda(\mathscr{N}_q)), \right. \\
 & \left. \min_{\mathscr{N}_c \in \mathscr{C}_q}
(B_2(\mathscr{N}_c) + 2 (\lambda(\mathscr{N}_q) - \lambda(\mathscr{N}_c))
\right\}. \\
\end{split}
\end{equation*}

Both $B_1(\mathscr{N}_q)$ and $B_2(\mathscr{N}_r)$ provide valid pruning rules.
We can combine both to get a tighter pruning rule by taking the tighter
of the two bounds.  In addition, $B_1(\mathscr{N}_q) \ge B_1(\mathscr{N}_c)$ and
$B_2(\mathscr{N}_q) \ge B_2(\mathscr{N}_c)$ for all $\mathscr{N}_c \in
\mathscr{C}_q$.  Therefore, we can prune $(\mathscr{N}_q, \mathscr{N}_r)$ if
$d_{\min}(\mathscr{N}_q, \mathscr{N}_r) \ge \min \{
B_1(\operatorname{Par}(\mathscr{N}_q)),
B_2(\operatorname{Par}(\mathscr{N}_q))\}.$

\newpage
These observations are combined for a better bound:

\vspace*{-1.3em}
\begin{equation*}
\begin{split}
B(\mathscr{N}_q) = \min &\left\{ \max \big\{ \max_{p \in \mathscr{P}_q}
D_p[k], \max_{\mathscr{N}_c \in \mathscr{C}_q} B(\mathscr{N}_c) \big\},
\right. \\
 & \left. \min_{p \in \mathscr{P}_q} \big(D_p[k] + \rho(\mathscr{N}_q) + \lambda(\mathscr{N}_q)\big)
,\right. \\
 & \left. \min_{\mathscr{N}_c \in \mathscr{C}_q} \Big(B(\mathscr{N}_c) + 2
\big(\lambda(\mathscr{N}_q) - \lambda(\mathscr{N}_c)\big)\Big) \right. \\
 & B(\operatorname{Par}(\mathscr{N}_q)) \bigg\} . \\
\end{split}
\end{equation*}
\vspace*{-0.3em}

As a result of this bound function being expressed recursively, previous bounds
can be cached and used to calculate the bound $B(\mathscr{N}_q)$ quickly.  We
can use this to structure \texttt{Score()} as given in Algorithm
\ref{alg:knn_score}.

\begin{algorithm}[tb]
  \begin{algorithmic}
    \STATE {\bfseries Input:} query node $\mathscr{N}_q$, reference node $\mathscr{N}_r$
    \STATE {\bfseries Output:} a score for the node combination $(\mathscr{N}_q,
\mathscr{N}_r)$, or $\infty$ if the combination should be pruned

    \medskip

    \IF{$d_{\min}(\mathscr{N}_q, \mathscr{N}_r) < B(\mathscr{N}_q)$}
      \RETURN $d_{\min}(\mathscr{N}_q, \mathscr{N}_r)$
    \ENDIF

    \medskip
    \RETURN $\infty$
  \end{algorithmic}

  \caption{$k$-nearest-neighbors \texttt{Score()}}
  \label{alg:knn_score}
\end{algorithm}

Applying the meta-algorithm in Section \ref{sec:overview} with any tree type and
any pruning dual-tree traversal gives a correct implementation of $k$-nearest
neighbor search.  Proving the correctness is straightforward; 
first, a (non-pruning) dual-tree traversal which uses \texttt{BaseCase()} as
given in Algorithm \ref{alg:knn_base_case} will give correct results for any
space tree.  Then, we already know that $B(\mathscr{N}_q)$ is a bounding
function that, at any point in the traversal, will not prune any subtrees which
could contain better nearest neighbor candidates than the current candidates.
Thus, the true nearest neighbors for each query point will always be visited,
and the results will be correct.

We now show that this algorithm is a generalization of the standard
$kd$-tree $k$-NN search, which uses a pruning dual-tree depth-first traversal.
The archetypal algorithm for all-nearest neighbor search ($k$-nearest neighbor
search with $k = 1$) given for $kd$-trees in Alex Gray's Ph.D. thesis
\yrcite{gray2003phd} is shown here in Algorithm \ref{alg:gray_knn} with
converted notation.  $\delta_q^{nn}$ is the bound for a node $\mathscr{N}_q$ and
is initialized to $\infty$; $D_{p_q}$ represents the nearest distance for a
query point $p_q$, and $N_{p_q}$ represents the nearest neighbor for a query
point $p_q$.  $\mathscr{N}_q$.left represents the left child of $\mathscr{N}_q$
and is defined to be $\mathscr{N}_q$ if $\mathscr{N}_q$ has no children;
$\mathscr{N}_q$.right is similarly defined.

The structure of the algorithm matches Algorithm \ref{alg:depth_traversal}; it
is a dual-tree depth-first recursion.  Because this is a depth-first recursion,
$\delta_q^{nn} = \infty$ for a node $\mathscr{N}_q$ if no descendants of
$\mathscr{N}_q$ have been recursed into.  Otherwise, $\delta_q^{nn}$ is the
maximum of $D_{p_q}$ for all $p_q \in \mathscr{D}^p_q$.  That is, $\delta_q^{nn}
= B_1(\mathscr{N}_q)$.  Thus, the comparison in the first line of Algorithm
\ref{alg:gray_knn} is equivalent to Algorithm \ref{alg:knn_score} with
$B_1(\mathscr{N}_q)$ instead of $B(\mathscr{N}_q)$.


The first two lines of the inside of the \textbf{for each} loop (the base case)
are equivalent to Algorithm \ref{alg:knn_base_case} with $k = 1$.  $kd$-trees
only hold points in leaves; therefore, the base case is called for all
combinations of points in each node combination, identically to the depth-first
traverser (Algorithm \ref{alg:depth_traversal}).

A $kd$-tree is a space tree and the dual depth-first recursion is a pruning
dual-tree traversal.  Also, we showed the equivalency of the pruning rule (that
is, the \texttt{Score()} function) and the equivalency of the base case.  So, it
is clear that Algorithm \ref{alg:gray_knn} is produced using our meta-algorithm
with these parameters.  In addition, because $B(\mathscr{N}_q)$ is always less
than $B_1(\mathscr{N}_q)$, Algorithm \ref{alg:knn_score} provides a {\bf tighter
bound} than the pruning rule in Algorithm \ref{alg:gray_knn}.

\begin{algorithm}[tb]
  \begin{algorithmic}
    \STATE \textbf{if} $d_{\min}(\mathscr{N}_q, \mathscr{N}_r) \ge
\delta_q^{nn}$, \textbf{then return}
    \IF{$\mathscr{N}_q$ is leaf \AND $\mathscr{N}_r$ is leaf}
      \FORALL{$p_q \in \mathscr{P}_q, p_r \in \mathscr{P}_r$}
        \STATE $d_{qr} \gets \| p_q - p_r \|$.
        \STATE \textbf{if} $d_{qr} < D_{p_q} $ \textbf{then} $D_{p_q} = d_{qr}$; $N_{p_q} = p_r$
        \STATE \textbf{if} $d_{qr} < \delta_q^{nn}$ \textbf{then} $\delta_q^{nn} \gets d_{qr}$
      \ENDFOR
    \ENDIF
    \STATE \texttt{AllNN($\mathscr{N}_q$.}left, closer-of$(\mathscr{N}_r$.left,
$\mathscr{N}_r$.right$)$\texttt{)}
    \STATE \texttt{AllNN($\mathscr{N}_q$.}left, farther-of$(\mathscr{N}_r$.left,
$\mathscr{N}_r$.right$)$\texttt{)}
    \STATE \texttt{AllNN($\mathscr{N}_q$.}right, closer-of$(\mathscr{N}_r$.left,
$\mathscr{N}_r$.right$)$\texttt{)}
    \STATE \texttt{AllNN($\mathscr{N}_q$.}right, farther-of$(\mathscr{N}_r$.left, $\mathscr{N}_r$.right$)$\texttt{)}
    \STATE $\delta_q^{nn} = \min(\delta_q^{nn},
\max(\delta_{q.\mathrm{left}}^{nn}, \delta_{q.\mathrm{right}}^{nn}))$
  \end{algorithmic}

  \caption{\texttt{AllNN($\mathscr{N}_q$, $\mathscr{N}_r$)} \cite{gray2003phd}}
  \label{alg:gray_knn}
\end{algorithm}

This algorithm is also a generalization of the standard cover tree $k$-NN search
\cite{langford2006}.  The cover tree search is a pruning dual-tree
traversal where the query tree is traversed depth-first while the reference
tree is simultaneously traversed breadth-first.  The pruning rule (after
simple adaptation to the $k$-nearest-neighbor search problem instead of the
nearest-neighbor search problem) is equivalent to

\vspace*{-0.8em}
$$ d_{\min}(\mathscr{N}_q, \mathscr{N}_r) \ge D_{p_q}[k] +
\lambda(\mathscr{N}_q) $$
\vspace*{-1.8em}

where $p_q$ is the point contained in $\mathscr{N}_q$ (remember, each node of a
cover tree contains one point).  This is equivalent to $B_2(\mathscr{N}_q)$
because $\rho(\mathscr{N}_q) = 0$ for cover trees.  The transformation from the
algorithm given by Beygelzimer et al.  \yrcite{langford2006} to our
representation is made clear in Appendix A (supplementary material) and in the
$k$-nearest neighbor search implementation of the C++ library MLPACK
\cite{curtin2011}; this is implemented in terms of our meta-algorithm.

\vspace*{-0.1em}
Specific algorithms for ball trees, metric trees, VP trees, octrees, and other
space trees are trivial to create using the \texttt{BaseCase()} and
\texttt{Score()} implementation given here (and in MLPACK).  Note also that this
implementation will work in any metric space.

\vspace*{-0.1em}
An extension to $k$-furthest neighbor search is straightforward.  The bound
function must be `inverted' by changing `max' to `min' (and vice versa); in
addition, the distances $D_{p_q}[i]$ must be initialized to $0$ instead of
$\infty$, and the lists $D$ and $N$ must be sorted by descending distance
instead of ascending distance.  Lastly, the comparison $d < D_{p_q}[k]$ must be
changed to $d > D_{p_q}[k]$.  With these simple changes, we have solved an
entirely different problem using our meta-algorithm with very little effort.  A
$k$-furthest neighbor search using our meta-algorithm for both $kd$-trees and
cover trees is also available in MLPACK.


\vspace*{-0.5em}
\section{Range Search}
\label{sec:range}

\vspace*{-0.1em}
Range search is another popular neighbor searching problem related to
$k$-nearest neighbor search.  In addition to being a fairly standard machine
learning task, it has numerous uses in applications such as databases and
geographic information systems (GIS).  A treatise on the history of the problem
and solutions is given by Agarwal \& Erickson \yrcite{agarwal1999}.  The problem
is:

\vspace*{-0.1em}
Given query and reference datasets $S_q, S_r$ and a range $[\delta_1,
\delta_2]$, for each point $p_q \in S_q$, find all points in $S_r$ such that
$\delta_1 \le \| p_q - p_r \| \le \delta_2$.  As with $k$-nearest neighbor
search, refer to the list of neighbors for each query point $p_q$ as $N_{p_q}$
and the corresponding distances as $D_{p_q}$.  These lists are not sorted in any
particular order, and at initialization time, they are empty.

\vspace*{-0.1em}
In different settings, the problem of range search may not be stated
identically; however, our results are easily adaptable.  A \texttt{BaseCase()}
implementation is given in Algorithm \ref{alg:rs_base_case}, and a
\texttt{Score()} implementation is given in Algorithm \ref{alg:rs_score}.  The
only bounds to consider are $[\delta_1, \delta_2]$, so no complex bound handling
is necessary.

\vspace*{-0.1em}
While range search is sometimes mentioned in the context of
dual-tree algorithms \cite{nbody}, the focus is usually on $k$-nearest neighbor
search.  As a result, we cannot find any explicitly published dual-tree
algorithms to generalize; however, a single-tree algorithm was proposed by
Bentley and Friedman \yrcite{bentley1979data}.  Thus, the \texttt{BaseCase()}
and \texttt{Score()} proposed here can be used with our meta-algorithm to
produce entirely novel range search implementations; MLPACK has $kd$-tree and
cover tree implementations.

\begin{algorithm}[tb]
  \begin{algorithmic}
    \STATE \textbf{Input:} query point $p_q$, reference point $p_r$, neighbor list
$N_{p_q}$, distance list $D_{p_q}$
    \STATE \textbf{Output:} distance $d$ between $p_q$ and $p_r$
    \medskip
    \STATE $ d \gets \| p_q - p_r \| $
    \medskip
     \IF{$\delta_1 \le d \le \delta_2 $ \AND \texttt{BaseCase($p_q$, $p_r$)} not
yet called}
       \STATE $N_{p_q} \leftarrow N_{p_q} \cup \{ p_r \}$
       \STATE $D_{p_q} \leftarrow D_{p_q} \cup \{ d \}$
     \ENDIF
     \RETURN $d$
  \end{algorithmic}
  \caption{Range search \texttt{BaseCase()}.}
  \label{alg:rs_base_case}
\end{algorithm}

\begin{algorithm}[tb]
  \begin{algorithmic}
    \STATE \textbf{Input:} query node $\mathscr{N}_q$, reference node
$\mathscr{N}_r$
    \STATE \textbf{Output:} a score for $(\mathscr{N}_q, \mathscr{N}_r)$, or
$\infty$ if the combination should be pruned
    \medskip
    \IF{$\delta_1 \le d_{min}(\mathscr{N}_q, \mathscr{N}_r) \le \delta_2 $}
      \RETURN $d_{min}(\mathscr{N}_q, \mathscr{N}_r)$
    \ENDIF
    \RETURN $\infty$
  \end{algorithmic}
  \caption{Range search \texttt{Score()}.}
  \label{alg:rs_score}
\end{algorithm}

\vspace*{-2em}
\section{Bor\r{u}vka's Algorithm}
\label{sec:dtb}


\vspace*{-0.3em}
Finding a Euclidean minimum spanning tree has been a relevant problem since
Bor\r{u}vka's algorithm was proposed in 1926.  Recently, a dual-tree
version of Bor\r{u}vka's algorithm was developed \cite{march2010} for $kd$-trees
and cover trees.  We unify these two algorithms and generalize to other types of
space tree by formulating \texttt{BaseCase()} and \texttt{Score()} functions.

\vspace*{-0.3em}
For a dataset $S_r \in \Re^{N \times D}$, Bor\r{u}vka's algorithm connects each
point to its nearest neighbor, giving many `components'.  For each component
$c$, the nearest point in $S_r$ to any point of $c$ that is not part of $c$ is
found.  The points are connected, combining those components.  This process
repeats until only one component---the minimum spanning tree---remains.

\vspace*{-0.3em}
During the algorithm, we maintain a list $F$ made up of $i$ components $F_i : \{
E_i, V_i \}$ where $E_i$ is the list of edges and $V_i$ is the list of vertices
in the component $F_i$ (these are points in $S_r$).  Each point in $S_r$ belongs
to only one $F_i$.  At initialization, $|F| = |S_r|$ and $F_i =
\{ \emptyset, \{ p_i \} \}$ for $i = \{ 1, \dots, |S_r| \}$, where $p_i$ is the
$i$'th point in $S_r$.  For $p \in S_r$ we define $F(p) = F_i$ if $F_i$ is the
component containing $p$.  During the algorithm, we maintain $N(F_i)$ as the
candidate nearest neighbor of component $F_i$ and $p_c(F_i)$ as the point in
component $F_i$ nearest to $N(F_i)$.  Then, $D(F_i) = \| p_c(F_i) - N(F_i) \|$.
Remember that $F(N(F_i)) \ne F_i$.

\begin{algorithm}[t!]
\begin{algorithmic}
    \STATE {\bfseries Input:} query point $p_q$, reference point $p_r$, nearest
candidate point $N(F(p_q))$ and distance $D(F(p_q))$
    \STATE {\bfseries Output:} distance $d$ between $p_q$ and $p_r$

    \medskip
    \IF{$p_q = p_r$}
      \RETURN $0$
    \ENDIF
    \IF{$F(p_q) \ne F(p_r)$ \AND $\| p_q - p_r \| < D(F(p_q))$}
      \STATE $D(F(p_q)) \gets \| p_q - p_r \|$
      \STATE $N(F(p_q)) \gets p_r$; $\;\;$ $p_c(F(p_q)) \gets p_q$
    \ENDIF
    \RETURN $\| p_q - p_r \|$
  \end{algorithmic}

  \caption{Bor\r{u}vka's algorithm \texttt{BaseCase()}.}
  \label{alg:dtb_base_case}
\end{algorithm}

\begin{algorithm}[t]
  \begin{algorithmic}
    \STATE {\bfseries Input:} query node $\mathscr{N}_q$, reference node $\mathscr{N}_r$
    \STATE {\bfseries Output:} a score for the node combination $(\mathscr{N}_q,
\mathscr{N}_r)$, or $\infty$ if the combination should be pruned

    \medskip

    \IF{$d_{\min}(\mathscr{N}_q, \mathscr{N}_r) < B(\mathscr{N}_q)$}
      \IF{$F(p_q) = F(p_r) \; \forall p_q \in \mathscr{D}^p_q, p_r \in
\mathscr{D}^p_r$}
        \RETURN $\infty$
      \ENDIF
      \RETURN $d_{\min}(\mathscr{N}_q, \mathscr{N}_r)$
    \ENDIF

    \RETURN $\infty$
  \end{algorithmic}

  \caption{Bor\r{u}vka's algorithm \texttt{Score()}.}
  \label{alg:dtb_score}
\end{algorithm}

\vspace*{-0.4em}
To run Bor\r{u}vka's algorithm with a space tree $\mathscr{T}_r$ built on the
set of points $S_r$, a pruning dual-tree traversal is run with
\texttt{BaseCase()} as Algorithm \ref{alg:dtb_base_case},
\texttt{Score()} as Algorithm \ref{alg:dtb_score}, $\mathscr{T}_r$ as
\textit{both} of the trees, and $F$ as initialized before.  Note that
\texttt{Score()} uses $B(\mathscr{N}_q)$ from Section \ref{sec:knn} with $k =
1$.  Upon traversal completion, we have a list $N(F_i)$ of nearest neighbors
of each component $F_i$.  The edge $(N(F_i), p_c(F_i))$ is added to $F_i$ for
each $F_i$.  Then, any components in $F$ with shared edges are merged into a new
list $F'$ where $|F'| < |F|$.  The pruning dual-tree traversal is then run again
with $F = F'$ and the traversal-merge process repeats until $|F| = 1$.  When
$|F| = 1$, then $F_1$ is the minimum spanning tree of $S_r$.

\vspace*{-0.4em}
To prove the correctness of the meta-algorithm, see Theorem 4.1 in March et~al.
\yrcite{march2010}.  That proof can be adapted from $kd$-trees to
general space trees.  Our representation is a generalization of their
algorithms; our meta-algorithm to produces their $kd$-tree and cover
tree implementations with a tighter distance bound $B(\mathscr{N}_q)$.  Our
meta-algorithm produces a provably correct dual-tree algorithm with any type of
space tree.

\vspace*{-0.7em}
\section{Kernel Density Estimation}
\label{sec:kde}

\vspace*{-0.3em}
Much work has been produced regarding the use of dual-tree algorithms
for kernel density estimation (KDE), including by Gray \& Moore \yrcite{nbody,
gray2003nonparametric} and later by Lee et al. \yrcite{lee2005, lee2008fast}.
KDE is an important machine learning problem with a vast range of applications,
such as signal processing to econometrics.  The
problem is, given query and reference sets $S_q, S_r$, to estimate a probability
density $f_q$ at each point $p_q \in S_q$ using each point $p_r \in S_r$ and a
kernel function $K_h$.  The exact probability density at a point $p_q$ is the
sum of $K(\| p_q - p_r \|)$ for all $p_r \in S_r$.

\begin{algorithm}[b]
  \begin{algorithmic}
    \STATE \textbf{Input:} query node $\mathscr{N}_q$, reference node
$\mathscr{N}_r$
    \STATE \textbf{Output:} a score for $(\mathscr{N}_q, \mathscr{N}_r)$ or
$\infty$ if the combination should be pruned
    \medskip
    \IF{$B_K(\mathscr{N}_q, \mathscr{N}_r) \ge \frac{\epsilon}{| S_r |}$}
      \FORALL{$p_q \in \mathscr{D}^p_q$}
        \STATE $f_q \gets f_q + |\mathscr{D}^p_r| K( \| p_q - C_r \| )$
      \ENDFOR
      \RETURN $\infty$
    \ENDIF
    \RETURN $d_{\min}(N_q, N_r)$
  \end{algorithmic}
  \caption{KDE \texttt{Score($\mathscr{N}_q$, $\mathscr{N}_r$)}.}
  \label{alg:kde_score}
\end{algorithm}

\begin{algorithm}[b!]
  \begin{algorithmic}
    \STATE \textbf{Input:} query point $p_q$, reference point $p_r$, density
estimate $f_q$
    \STATE \textbf{Output:} distance between $p_q$ and $p_r$
    \medskip
    \STATE \textbf{if} \texttt{BaseCase($p_q$, $p_r$)} already called
\textbf{then return}
    \STATE $f_q \leftarrow f_q + K_h( \| p_q - p_r \| )$
    \RETURN $\| p_q - p_r \|$
  \end{algorithmic}
  \caption{KDE \texttt{BaseCase($p_q$, $p_r$)}.}
  \label{alg:kde_base_case}
\end{algorithm}

In general, the kernel function is some zero-centered probability density
function, such as a Gaussian.  This means that when $\| p_q - p_r \|$ is very
large, the contribution of $K$ to $f_q$ is very small.  Therefore, we can
approximate small values using a dual-tree algorithm to avoid unnecessary
computation; this is the idea set forth by Gray \& Moore \yrcite{nbody}.
Because $K$ is a function which is decreasing with distance, the maximum
difference between $K$ values for a given combination $(\mathscr{N}_q,
\mathscr{N}_r)$ can be bounded above with

\vspace*{-2.3em}
$$ B_K(\mathscr{N}_q, \mathscr{N}_r) = K(d_{\min}(\mathscr{N}_q, \mathscr{N}_r))
- K(d_{\max}(\mathscr{N}_q, \mathscr{N}_r)). $$
\vspace*{-2.3em}

The algorithm takes a parameter $\epsilon$; when $ B_K(\mathscr{N}_q,
\mathscr{N}_r) $ is less than $\epsilon / | S_r |$, the kernel
values are approximated using the kernel value of the centroid $C_r$ of the
reference node.  The division by $| S_r |$ ensures that the total approximation
error is bounded above by $\epsilon$.  The base case on $p_q$ and $p_r$ merely
needs to add $K( \| p_q - p_r \| )$ to the existing density estimate $f_q$.
When the algorithm is initialized, $f_q = 0$ for all query points.
\texttt{BaseCase()} is Algorithm \ref{alg:kde_base_case} and
\texttt{Score()} is Algorithm \ref{alg:kde_score}.


\vspace*{-0.3em}
Again we emphasize the flexibility of our meta-algorithm.  To our knowledge
cover trees, octrees, and ball trees have never been used to perform
KDE in this manner.  Our meta-algorithm can produce these implementations with
ease.


\vspace*{-1.1em}
\section{Discussions}
\label{sec:discussion}

\vspace*{-0.2em}
We have now shown five separate algorithms for which we have taken existing
dual-tree algorithms and constructed a \texttt{BaseCase()} and \texttt{Score()}
function that can be used with any space tree and any dual-tree traversal.
Single-tree extensions of these four methods are straightforward
simplifications.

This modular way of viewing tree-based algorithms has several useful immediate
applications.  The first is implementation.  Given a tree implementation and a
dual-tree traversal implementation, all that is required is \texttt{BaseCase()}
and \texttt{Score()} functions.  Thus, code reuse can be maximized, and new
algorithms can be implemented simply by writing two new functions.  More
importantly, the code is now modular.  MLPACK \cite{curtin2011}, written in C++,
uses templates for this.  One example is the \texttt{DualTreeBoruvka} class,
which implements the meta-algorithm discussed in Section \ref{sec:dtb}, and has
the following arguments:

\vspace*{-0.7em}
\begin{verbatim}
    template<typename MetricType,
             typename TreeType,
             typename TraversalType>
    class DualTreeBoruvka;
\end{verbatim}
\vspace*{-0.6em}

This means that any class satisfying the constraints of the \texttt{TreeType}
template parameter can be designed without any consideration or knowledge of the
\texttt{DualTreeBoruvka} class or of the \texttt{TraversalType} class; it is
\textit{entirely independent}.  Then, assuming a \texttt{TreeType} and
\texttt{TraversalType} without bugs, the dual-tree Bor\r{u}vka's algorithm is
guaranteed to work.  An immediate example of the advantage of this is that cover
trees were implemented for MLPACK for use with $k$-nearest neighbor search.
This cover tree implementation could, without any additional work, be used with
\texttt{DualTreeBoruvka}---which was never an intended goal during the cover
tree implementation but still a particularly valuable result!






\vspace*{-0.2em}
Of course, the utility of these abstractions are not limited implementation
details.  Each of the papers cited in the previous sections describe algorithms
in terms of one specific tree structure.  March et~al.  \yrcite{march2010}
discuss implementations of Bor\r{u}vka's Algorithm on both $kd$-trees and cover
trees and give algorithms for both.  Each algorithm given is quite different and
it is not easy to see their similarities.  Using our meta-algorithm, any of
these tree-based algorithms can be expressed with less effort---especially for
more complex trees like the cover tree---and in a more general sense.

\vspace*{-0.2em}
In addition, correctness proofs for our algorithms tend to be quite simple.  The
proofs for each algorithm here can be given in two simple sub-proofs:
\textit{(1)} prove the correctness of \texttt{BaseCase()} when no prunes are
made, and \textit{(2)} prove that \texttt{Score()} does not prune any subtrees
which the correctness of the results depends on.

\vspace*{-0.2em}
The logical split of base case, pruning rule, tree type, and traversal can also
be advantageous.  A strong example of this is the function $B(\mathscr{N}_q)$
devised in Section \ref{sec:knn}, which is a novel, tighter bound.  When not
considering a particular tree, the path to a superior algorithm can often be
simpler (as in that case).

\subsection{Parallelism}

Nowhere in this paper has parallelism been discussed in any detail.  In fact,
all of the given algorithms seem to be suited to serial implementation.
However, the pruning dual-tree traversal is entirely separate from the rest of
the dual-tree algorithm; therefore, a parallel pruning dual-tree traversal can
be used without modifying the rest of the algorithm.

\vspace*{-0.2em}
For instance, consider $k$-nearest neighbor search.  Most large-scale parallel
implementations of $k$-NN do not use space trees but instead techniques like
LSH for fast (but inexact) search.  To our knowledge, no freely available
software exists that implements distributed dual-tree $k$-nearest neighbor
search.

\vspace*{-0.2em}
As a simple (and not necessarily efficient) proof-of-concept idea for a
distributed traversal, suppose we have $t^2$ machines and a ``master'' machine
for some $t > 0$.  Then, for a query tree $\mathscr{T}_q$ and a reference tree
$\mathscr{T}_r$, we can split $\mathscr{T}_q$ into $t$ subtrees and one ``master
tree'' $\mathscr{T}_{qm}$.  The reference tree $\mathscr{T}_r$ is split the same
way.  Each possible combination of query and reference subtrees is stored on one
of the $t^2$ machines, and the master trees are stored on the master machine.
The lists $D$ and $N$ can be stored on the master machine and can be updated or
queried by other machines.

\vspace*{-0.2em}
The traversal starts at the roots of the query tree and reference tree and
proceeds serially on the master.  When a combination in two subtrees is reached,
\texttt{Score()} and \texttt{BaseCase()} are performed on the machine containing
those two subtrees and that subtree traversal continues in parallel.  This idea
satisfies the conditions of a pruning dual-tree traversal; thus, we can use it
to make any dual-tree algorithm parallel.

\vspace*{-0.2em}
Recently, a distributed dual-tree algorithm was developed for
kernel summation \cite{lee2012distributed}; this work could be adapted to a
generalized distributed pruning dual-tree traversal for use with our
meta-algorithm.

\vspace*{-0.5em}
\section{Conclusion}


\vspace*{-0.2em}
We have proposed a tree-independent representation of dual-tree algorithms and a
meta-algorithm which can be used to create these algorithms.  A dual-tree
algorithm is represented as four parts: a space tree, a pruning dual-tree
traversal, a \texttt{BaseCase()} function, and a \texttt{Score()} function.  We
applied this representation to generalize and extend five example dual-tree
algorithms to different types of trees and traversals.  During this process, we
also devised a novel bound for $k$-nearest neighbor search that is tighter than
existing bounds.  Importantly, this abstraction can be applied to help approach
the problem of parallel dual-tree algorithms, which currently is not well
researched.

\bibliography{paper}
\bibliographystyle{icml2013}

\end{document}